\documentclass[9pt,a4paper,twoside]{tau-class/tau}
\usepackage[english]{babel}

\journalname{Physics in Silico}
\title{Astrophysical Narratives: Poetic Representations of Gamma-Ray Emission from FermiLAT via Markov Chains}

\author[a]{Carlos Darío Badilla Cerdas }

\affil[a]{Escuela de Física, Universidad de Costa Rica}
\affil{\url{carlos.badillacerdas@ucr.ac.cr}}

\institution{Universidad de Costa Rica}
\footinfo{|PhiS$\rangle$}

\begin{abstract}    
The intersection of art and science offers novel ways to interpret and represent complex phenomena. This project explores the convergence of high-energy astrophysics, concrete poetry, and natural language processing (NLP) by proposing a Markov chain-based algorithm that generates poetic texts from gamma-ray emission maps of the universe. Gamma rays, the most energetic form of electromagnetic radiation, arise from violent astrophysical processes such as supernovae, pulsars, and black hole accretion, observable through instruments like the \textit{Fermi Large Area Telescope} (FermiLAT). These high-energy events are mapped and processed into matrices, treated as Markov chains, where each state's transition probability is determined by the intensity of gamma-ray sources in a region of interest.
\end{abstract}


\begin{document}
		
    \maketitle 
    \thispagestyle{firststyle} \tauabstract

\section{The First Spark}

For millennia, humanity has turned its gaze to the night sky, seeking in the stars both inspiration and answers to the mysteries of the universe. Artists, in particular, have found in the heavens an inexhaustible source of creation. From masterpieces like Vincent van Gogh’s \textit{Starry Night} \cite{vangogh} to lesser-known works, the sky has served as a canvas for expressing humanity's deepest emotions. Poets especially have seen in the constellations a cosmic narrative where fundamental human concepts—love and its absence, melancholy, happiness, grandeur, gods, and demons—find a sanctuary \cite{davidlevy}. In the vast firmament, everything that dwells in the human imagination seems to come to life.

As we move beyond our atmosphere, exploring the confines of the universe with modern instruments, we encounter physical processes that surpass any human narrative. The birth and death of stars, black holes that consume every light, supernovae marking the end of massive stars, and cosmic dances of massive objects distorting space-time itself. These phenomena, of almost inconceivable magnitude, have fascinated scientists and poets alike, creating a bridge between astrophysical observation and artistic expression. The aim of this project is precisely that: to intertwine poetry and science, granting a literary voice to the most violent astrophysical events in the universe.

In particular, high-energy astrophysics, dedicated to studying the processes that generate the highest amounts of energy, provides us with a unique window to the cosmos. Gamma-ray emissions, the most energetic particles of light, are a crucial means for observing and studying events such as blazars, pulsars, and quasars. These phenomena, though distant and abstract, become poetic narrators in this work.

To accomplish this task, an algorithm based on Markov chains is proposed for text generation. This algorithm uses a gamma-ray source map of a specific sky region of interest (ROI), captured by the \textit{Fermi Large Area Telescope} (FermiLAT), as the transition matrix. Thus, each word of the generated poem is influenced by the energetic activity of astronomical objects, creating a symbiosis between astrophysical data and human language.

This idea arises from a reflection on the intersection of science and art. Jim Jarmusch’s film \textit{Paterson} \cite{paterson}, a deeply introspective work on poetry, was a key source of inspiration for the development of this algorithm. In \textit{Paterson}, the protagonist finds beauty and meaning in the smallest details of everyday life, using language as a medium to give them life and transcendence. Similarly, this work aims to give a poetic voice to cosmic events through the lens of human emotions, much like Paterson transforms the mundane into poetry.

Moreover, this project emerges not only from a creative quest but also from a personal need. Throughout history, art has been an essential refuge for those facing emotional and mental challenges. An iconic example is Vincent van Gogh, who, despite his struggles with bipolar disorder and depression \cite{blumer2002illness, wolf2001creativity}, found in art a way to channel his emotions and stay connected to the world. Personally, poetry has been a crucial medium for maintaining that same connection. During my stay in a mental health hospital, poetic creation allowed me to reconnect both with my surroundings and with myself, serving as a bridge between my inner world and the external reality.

Thus, this project becomes not only a scientific and literary exploration but also a manifestation of how art can transform and give meaning to the human experience, even in its darkest moments. In this sense, the combination of poetry and science seeks not only to capture the essence of astrophysical phenomena but also to give a delicate and emotive voice to the most violent and distant processes of the universe, connecting them to human emotions and experiences.

\section{Understanding the Universe: The Dance of Photons}

To comprehend the functioning and background of the algorithm, it is necessary to keep in mind three areas of knowledge: poetry, high-energy astrophysics, and Markov chains. Some concepts of interest from each area are described below.

\subsection{On Poetry}
Poetry is one of the oldest forms of artistic expression, where language is structured creatively and emotively. Throughout history, it has evolved into various styles and forms, each with its own distinctive characteristics. This project focuses 
particularly in two movements: free verse poetry and concrete poetry.

Free verse poetry is characterized by its lack of metric and rhyme restrictions, allowing content and form to align more freely and naturally. This approach invites the poet to focus on the meaning and emotionality of the words, liberating creation from the rigidity of traditional forms. In this sense, subjective expression is prioritized, where rhythm, imagery, and metaphors can flow without the limitations imposed by formal structure \cite{strachan2011poetry}. This style becomes a space where the author's soul manifests fully, reflecting the complexities of human experience.

On the other hand, concrete poetry, which emerged in the 1950s (though examples can be found as early as the 19th century), proposes a visual approach to language, where typographical form and design play a crucial role in the interpretation of the poem \cite{strachan2011poetry}. In concrete poetry, the arrangement of words on the page is not just a means of conveying a message but becomes a component of the meaning itself. This form of poetry utilizes space, typography, and colors to create visual experiences that complement the verbal dimension of the text. The interaction between form and content allows readers to explore new dimensions of meaning, creating a dialogue between language and its visual representation.

In the current era, the ease of computers allows concrete poetry to reach extremely high levels of abstraction, alongside tools such as \texttt{generativepoetry-py} \cite{generative}, \texttt{pronouncingpy} \cite{generative2}, and \texttt{olipy} \cite{generative3}. These Python libraries not only facilitate the creation of dynamic poetic works but also empower artists to experiment with nonlinear and visually complex structures. The convergence of these tools not only enhances poetic creation but also invites artists to reflect on the role of the author in the creative process. In this context, poetry becomes a space for dialogue between humans and machines, where the boundaries between creativity and algorithmicity blur, opening up a range of possibilities for exploring new forms of expression. 

\begin{center}
This license certifies \\ 
That \textit{the universe} \footnote{This is a modification of the poem Poetic License written in 1976 by Ron Padgett. The name \textit{Ron Padgett} was changed to \textit{the universe}.} may \\
tell whatever lies \\
His heart desires \\
Until it expires \cite{lehman2006oxford}\\
\end{center}

Furthermore, the New York School of Poetry, active primarily in the mid-20th century, has greatly influenced the development of contemporary poetry. This movement, which includes figures like Frank O'Hara, John Ashbery, and Ron Padgett (from whom the poems presented in Paterson are taken), is characterized by its focus on everyday experience and the use of colloquial and accessible language. New York poetry moves away from lyrical traditions and ventures into the realm of spontaneity, where the poem becomes a reflection of urban life and social interactions. The poets of this school often employ a blend of genres and styles, as well as a freer structure that allows for the inclusion of fragments of conversations, cultural references, and a sense of irreverent humor.

\subsection{On Gamma Rays and High-Energy Astrophysics}

Gamma rays are the most energetic form of electromagnetic radiation. They are characterized by the highest frequencies and, consequently, the shortest wavelengths in the electromagnetic spectrum, with energies exceeding 100 keV. This radiation is produced in some of the most extreme events in the universe, from supernova explosions to the violent activity of supermassive black holes \cite{NASA_pro2}. In terms of energy, the relationship between the energy of a gamma-ray photon \( E \) and its frequency \( \nu \) is given by Planck's equation \cite{eisberg1986quantum}:

\begin{equation}
E = h \nu
\end{equation}

where \( h \) is Planck's constant (\(6.626 \times 10^{-34}\, \text{Js}\)).
Gamma rays are fundamental to high-energy astrophysics because they reveal events and processes that cannot be observed at other wavelengths. These emissions are often associated with non-thermal phenomena, where charged particles (such as electrons) are accelerated to relativistic speeds, producing high-energy electromagnetic radiation.

\subsubsection{Gamma-Ray Emission Mechanisms}

Several astrophysical processes are responsible for the production of gamma rays. The main emission mechanisms include synchrotron radiation, inverse Compton scattering, electron-positron pair annihilation, and radioactive decay \cite{NASA_pro}.

\paragraph{Synchrotron Radiation:}

Synchrotron radiation is generated when charged particles, such as electrons, move at relativistic speeds in the presence of a magnetic field \cite{longair2011high}. Under these conditions, particles follow curved paths due to the Lorentz force, inducing the emission of electromagnetic radiation, predominantly in the radio range, although it can extend to gamma rays in highly energetic environments with intense magnetic fields. The total power \( P \) emitted by a charged particle moving in a magnetic field \( B \) at an isotropic velocity \cite{rybicki1991radiative} close to the speed of light \( v \approx c \) is described by the equation \cite{blumenthal1970bremsstrahlung}:

\begin{equation}
P = \frac{2 e^4 B^2 \gamma^2}{3 m_e^2 c^3}
\end{equation}

where \( e \) is the electron charge, \( \gamma \) is the Lorentz factor characterizing relativistic effects, \( m_e \) is the electron mass, and \( c \) is the speed of light in a vacuum. In astrophysical contexts, such as active galactic nuclei \cite{la2017relativistic} and supernova remnants, this radiation can reach gamma-ray energies if the magnetic field is sufficiently intense and particles reach high relativistic speeds.

\paragraph{Inverse Compton Process:}

The inverse Compton process occurs when charged particles, typically electrons, scatter low-energy photons (such as those from the microwave background radiation \cite{sunyaev1980microwave, zeldovich1969interaction}), transferring part of their kinetic energy to the photons. This increases the energy of the scattered photon, which can reach the gamma-ray region in high-energy conditions \cite{longair2011high}. The maximum energy \( E' \) that the photon can reach after scattering, for an electron with relativistic energy \( E_e \), is given by \cite{blumenthal1970bremsstrahlung}:

\begin{equation}
E' \approx 4\gamma^2 E_{\gamma}
\end{equation}

where \( \gamma \) is the electron's Lorentz factor and \( E_{\gamma} \) is the initial energy of the photon before scattering. This mechanism is fundamental in generating gamma rays in various astrophysical sources, such as active galactic nuclei and supernova remnants \cite{dermer2009high}, where high-energy photons are observed as a result of interactions between relativistic electrons and low-energy photons \cite{schlickeiser1979astrophysical}.

\paragraph{Electron-Positron Pair Annihilation:}

Positrons are created abundantly in the galaxy, and this abundance leads to one of the major energy loss processes for electrons: electron-positron pair annihilation \cite{guessoum2005lives}. This occurs when an electron \( e^- \) and a positron \( e^+ \) collide and annihilate each other, generating two gamma-ray photons, each with an energy of 511 keV, corresponding to the rest mass energy of both particles. This process is described by the reaction \cite{dermer2009high}:

\begin{equation}
e^- + e^+ \rightarrow \gamma + \gamma
\end{equation}

In high-energy conditions, such as near black holes and neutron stars \cite{dermer2009high, longair2011high}, where the densities of relativistic particles are significant, the probability of these interactions increases. Additionally, there are various sources of positrons in astronomical environments; one of the most common is the decay of positively charged pions (\( \pi^+ \)), produced in collisions between cosmic ray protons and interstellar gas. Positrons can also be generated via the decay of radioactive isotopes \cite{siegert2023positron}, such as \( ^{26} \text{Al} \), originating in supernova explosions and the decay of neutral pions into electron-positron pairs (\( \pi^0 \rightarrow e^- + e^+\)) \cite{husek2024neutral}. Electron-positron pair creation can occur through high-energy photon collisions with a nucleus field or even through photon-photon collisions in high-luminosity regions \cite{ruffini2010electron}. In photon-photon collisions, the threshold energy for pair production depends on the energy of the incident photons and their relative angle, reaching minimum values in head-on collisions. This phenomenon is particularly relevant near active galactic nuclei and high-energy X-ray sources, where photons generated at energies above this threshold tend to interact and may be absorbed, contributing to the opacity of the region to gamma rays.
Moreover, electrons and positrons can form positronium atoms \cite{dermer2009high}, a short-lived bound state that can decay into two 511 keV photons (in the singlet state) or three photons (in the triplet state), generating a continuous spectrum on the lower-energy side of the 511 keV line.

\paragraph{Radioactive Decay:}

Nuclear reactions occurring in supernova explosions and other extreme astrophysical environments can produce unstable radioactive nuclei, which emit gamma rays upon decay. A well-known example is the decay of \( ^{26}\text{Al} \) in stellar nursery regions, which, in addition to the aforementioned positron emission, emits gamma rays \cite{diehl2006radioactive}. 

None of the gamma-ray sources mentioned operate in isolation. It is common for gamma-ray emission to trigger a chain of events, including both radiation emission and absorption, as well as sequences of particle creation and annihilation.

\subsubsection{Gamma-Ray Observation with the Fermi LAT}

The \textit{Fermi Large Area Telescope} (Fermi LAT), part of the Fermi Gamma-ray Space Telescope \cite{fermi1}, is one of the most advanced instruments for gamma-ray observation. Launched in 2008, it has provided an unprecedented window into the universe at energies above 100 MeV. Due to the very nature of gamma rays, they cannot be measured directly; instead, the Fermi LAT detects gamma rays through their interaction with materials in the detector, producing electron-positron pairs that can then be tracked and analyzed \cite{atwood2009large}. One of the key outcomes of the Fermi LAT observations is the production of TS (Test Statistic) residual maps. These are tools used to assess the presence of signals amid background noise. These maps are constructed by comparing the likelihood of a source being present at a specific position with the null hypothesis, that is, the absence of a significant signal \cite{mattox1996likelihood}. The TS statistic is used to measure the difference between these models, assigning high TS values to regions where a gamma-ray source is more likely to exist. Generally, a TS value greater than 25 indicates a significant detection with a probability of around 5 sigma, while lower values reflect areas dominated by noise or weak sources. TS maps are especially useful in environments with multiple sources or in regions where complex background variations are expected, as they enable the localization and characterization of gamma-ray sources with greater precision.

\subsection{On Markov Chains}

A Markov chain is a mathematical model that describes a stochastic process in which the probability of an event occurring depends solely on the current state of the system, without considering prior states. This characteristic, known as the Markov property, is formally expressed as \cite{durrett2019probability}:

\begin{align}
\begin{split}
P(X_{n+1} = x \mid X_n = x_n, X_{n-1} = x_{n-1}, \dots, X_0 = x_0) &= \\
P(X_{n+1} = x \mid X_n = x_n)
\end{split}
\end{align}

where \( X_n \) represents the state of the system at time \( n \). This equation indicates that the probability of transition to the next state \( X_{n+1} \) depends solely on the current state \( X_n \), disregarding any prior states.

\subsection{States and Transitions}

In a discrete-time Markov chain, a set of \textbf{states} \( S = \{ s_1, s_2, \dots, s_k \} \) is defined along with a \textbf{transition matrix} \( P \) of dimension \( k \times k \), where each element \( P_{ij} \) represents the probability of transitioning from state \( i \) to state \( j \) in a single step. Formally, it is defined as \cite{norris1998markov}:
\begin{equation}
P_{ij} = P(X_{n+1} = s_j \mid X_n = s_i)
\end{equation}

The matrix \( P \) satisfies that each row sums to 1, meaning:

\begin{equation}
\sum_{j=1}^k P_{ij} = 1 \quad \forall i \in \{1, \dots, k\}
\end{equation}

This ensures that the probability of transition from any state \( i \) to all possible states forms a valid probability distribution.

\section{The Spectrum of Memory: The Data}

\subsection{Text Corpus}

The first type of data is the corpus, consisting of a selection of texts that will serve as input for the algorithm. Although texts of any type can be used, due to the nature of the algorithm, poetry and song lyrics are recommended. The larger the corpus, the better the algorithm's results may be, as greater lexical and structural diversity provides a richer context for poetic generation. For optimal algorithm performance, basic preprocessing of the corpus is advised, starting with tokenization by words, followed by lemmatization. Tokenization divides the text into manageable units \cite{webster1992tokenization}, while lemmatization converts each word to its base form \cite{plisson2004rule,jurafskyspeech}, facilitating linguistic analysis by removing the interference of conjugations or variations. This approach generalizes the corpus, eliminating specific conjugations while retaining the most general sense of each word.

Humans do not create from nothing; to enrich our knowledge, we need to consume external material. Whether in art or science, we learn primarily through imitation. Each author is partly shaped by everything they have encountered over their lifetime. Every artist they have seen or read becomes, unconsciously, part of that author. For this reason, incorporating small percentages of texts outside the author’s own work is recommended to enhance the fluency and breadth of the generation algorithm.

\subsection{Gamma-Ray Source Maps}

The second type of data consists of gamma-ray source maps, obtained by processing FermiLAT data. For our project, maps produced through a quick analysis using the \texttt{easyfermi} tool \cite{de2022easyfermi, wood2017fermipy, donath2023gammapy, price2018astropy, foreman2013emcee, saldana2021observational} are sufficient. However, for a higher level of detail, results from more exhaustive analyses may be used. The generated maps are in \texttt{FITS} format (Flexible Image Transport System), which is a standard in astronomy for storing and transmitting image and spectral data. They can be used directly in this format with the appropriate libraries, such as Astropy \cite{price2018astropy}, or quickly converted to more standard formats, like \texttt{PNG}, using tools such as \texttt{SAOImage DS9} \cite{sao}. For the selected map to effectively serve this purpose, a slight modification is necessary: row normalization must be applied. This ensures that the transition matrix satisfies the property in which the sum of transition probabilities from a given state equals one $( \sum_{j} P(i \to j) = 1 \quad \forall i)$.

\begin{figure}
    \centering
    \includegraphics[width=8 cm]{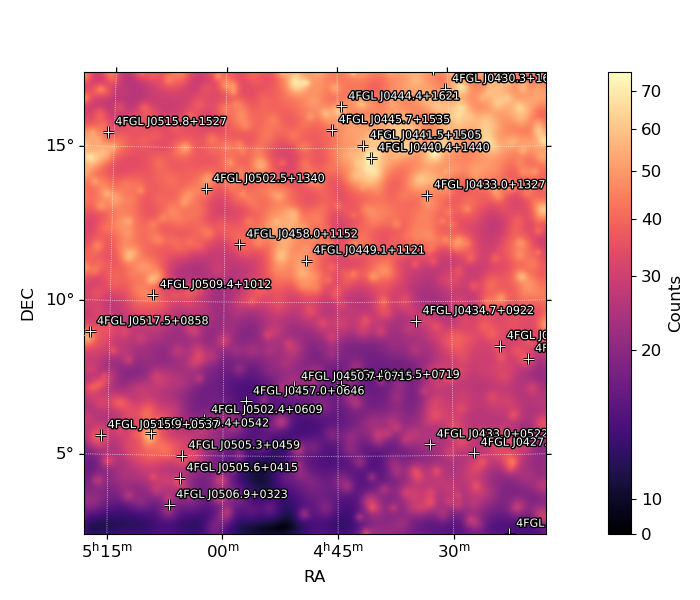}
    \caption{Statistical map of the ROI centered at coordinates RA 72.00 and Dec 9.9799°.}
    \label{fig:1}
\end{figure}

Figure \ref{fig:1} presents the map of the ROI used, showing gamma-ray sources cataloged in the latest Fermi catalog, 4FGL-DR4 \cite{ballet2023fermi}, within its boundaries.

\section{The Architecture of the Poem: Algorithm Description}

The proposed algorithm consists of a modified linguistic generation algorithm based on Markov chains. For its application, it is necessary to prepare a corpus to feed it and a gamma-ray source map that will serve as the probability distribution for the algorithm. The algorithm steps are described as follows:

\begin{enumerate}
    \item \textbf{Read the corpus:} A dictionary is created that associates each word in the corpus with its immediate consecutive words. For each word analyzed, the following words are noted, avoiding repetitions. This is fundamental, as the weight of each word will later be determined by gamma rays.
    
    \item \textbf{Create the poem:} Once the dictionary is constructed, a word is randomly chosen to begin text generation. Using the Markov process, the current state is determined through all possible following words, assigning a uniform probability to each one.
    
    \item \textbf{Define the Markov state matrix:} This matrix is determined from the selected gamma-ray source map. In applying the Markov process, the word with the highest value, associated with the most energetic region of the map, is selected as the next word.
    
    \item \textbf{Iterate the process:} The selected word becomes the current word, and the process is repeated until the desired maximum word count is reached.
    
    \item \textbf{Apply text correction:} Upon completion of the poem generation, a final filter consisting of text correction is applied, which can be carried out manually or using a language model (LLM). \label{item4}
\end{enumerate}

Since the algorithm is ``naive," the generated text lacks form and logic. This means it is not capable of creating proper conjugations. Spanish is a complex language in terms of its verb conjugations, with 16 different verb tenses that must be conjugated across 8 subject pronouns. This complexity exceeds the scope of the Markov analysis used and cannot be adequately handled in this implementation. To address this situation, an additional layer is added to the algorithm, which is an LLM. Current language models allow for tasks such as text correction to be performed quickly and efficiently. It is recommended to use this option in combination with manual supervision. It is important to review the raw results generated by the Markov algorithm to validate the LLM’s correct functioning.

\subsection{Mathematical Description of the Algorithm}
To facilitate its computational implementation, the algorithm is described as follows:

\begin{enumerate}
    \item \textbf{Read the corpus:}
    
    We define the corpus \( C = \{w_1, w_2, \dots, w_n\} \), where each \( w_i \) represents a word in the corpus. We construct a transition dictionary \( D \) such that for each word \( w_i \), the associated set \( S(w_i) \) of consecutive words in the corpus is:
    \[
    D(w_i) = \{w_j \mid w_j \in C\}.
    \]
    where \( w_j \) is the following word of \( w_i \). This dictionary excludes repetitions to capture only the first occurrence of each direct transition.

    \item \textbf{Define the Markov state matrix:}
    
    We define a transition matrix \( \mathbb{M} \) of size \( |C| \times |C| \), where each element \( m_{ij} \) represents the transition probability from \( w_i \) to \( w_j \), determined based on the gamma energy map:
    \[
    m_{ij} = \frac{\gamma_{ij}}{\sum_{k} \gamma_{ik}},
    \]
    where \( \gamma_{ij} \) is the gamma-ray energy value associated with the transition \( w_i \to w_j \).

    \item \textbf{Create the poem:}
    
    An initial word \( w_0 \in C \) is chosen randomly. Then, using the Markov model, we generate the word sequence by selecting the next word \( w_{i+1} \) according to the set \( S(w_i) \) of possible next words with uniform probability:
    \[
    P(w_{i+1} \mid w_i) = \frac{S(w_i)}{|S(w_i)|} \cdot \mathbb{M} \quad \forall w_{i+1} \in S(w_i).
    \]

    \item \textbf{Iterate the process:}
    
    Starting from the initial word \( w_0 \), we iterate until reaching the desired number of words \( m \). In each step \( i \), we select \( w_{i+1} \) using the transition value of \( \mathbb{M} \) corresponding to the maximum energy:
    \[
    w_{i+1} = \arg\max_{w_j \in S(w_i)} m_{ij}.
    \]

    \item \textbf{Apply text correction:}
    
    Upon completion of generation, a final correction \( T \) is applied to the resulting poem, using either a language model or manual review, ensuring that the text has adequate coherence and grammar:
    \[
    T(\text{poem}) \to \text{corrected poem}.
    \]
\end{enumerate}

\section{The Crystallization of the Poem: Implementation}

The proposed algorithm was implemented in Python due to its versatility, extensive NLP library functions, and its capabilities for integrating artificial intelligence into workflows. This section describes the specific steps of the implementation, from corpus construction to final processing with language models, highlighting the decisions and specific adjustments for using astrophysical data as the probability distribution in the generative process.

\subsection{Corpus Construction}

For this work, a custom \textit{corpus} was employed, composed of poems, stories, and personal letters from the author, supplemented with a selection of popular songs. It is important to remember that the algorithm is limited and only works by extracting probabilities from the provided \textit{corpus}. This \textit{corpus} was processed in Python using tokenization and lemmatization functions, allowing both the word order and punctuation marks to be preserved.

\subsection{Gamma-Ray Based Probability Distribution}

\begin{figure}
    \centering
    \includegraphics[width=8cm]{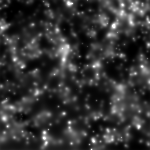}
    \caption{Final map used as the model's transition matrix, centered on coordinates RA 72.00 and Dec 9.9799°.}
    \label{fig:2}
\end{figure}

The Markov transition matrix was constructed using a gamma-ray source map, centered on astronomical coordinates with right ascension (\textit{RA}) of $4\,\text{h}\,48\,\text{m}$ and declination (\textit{Dec}) of $9.9799^\circ$, covering the time period from February 17, 2020, to March 8, 2020. The choice of these coordinates and dates is based on a personal criterion of the author, who seeks to imbue the generated poems with a specific emotional charge, ideal for poetic creation based on deep feelings—(what can I say? I'm a hopeless romantic). Figure \ref{fig:2} presents the final map used, with minor color distribution modifications made using \texttt{SAOImage DS9} to highlight point sources.

\subsection{Dimensional Adjustment of the State Matrix}

The dimensions of the state matrix may vary relative to the corpus word vector, leading to two potential problematic situations: (1) when the number of words exceeds the matrix dimension, and (2) when the number of words is less than the matrix dimension. To address these issues and ensure proper alignment, an image scaling algorithm was used to adjust the matrix to the required dimensions without losing the initial probabilistic structure. This procedure was performed through interpolation, preserving the gamma-ray density distribution in each transformation.

\subsection{Spell Check and Poetic Generation}

To streamline the final correction step, Python spell-checking libraries were implemented. These orthographic functions reduce basic errors in the text generation output, offloading part of the correction work to the LLM used subsequently. Additionally, some supplementary functions were developed to enhance the poetic quality of the generated text:

\begin{itemize}
    \item \textbf{Synonyms to Enrich Vocabulary:} A natural limitation of algorithms based on small corpora is the restriction in available vocabulary. To address this, a function was implemented that searches for synonyms of selected words when the transition probability of a word exceeds a certain threshold. Words associated with regions of higher gamma-ray density have a greater likelihood of being replaced by synonyms, enriching the text and expanding the algorithm's expressiveness without deviating from the original content of the corpus.

    \item \textbf{Free Verse Formatting:} Most traditional poems are organized in short lines of verse that emphasize specific words or phrases. To emulate this structure in the generated text, a function was incorporated that adds line breaks based on the probability of each word, as determined by the gamma-ray density in the corresponding region of the map. This free verse structure allows for the creation of lines that may consist of a single word or even a number, aligning with the tradition of contemporary poetry without metric restrictions.
\end{itemize}

\subsection{Final Correction with a Natural Language Model}

For the final correction of the generated poems, a pre-trained LLM operated locally was used. Among the available options, the GPT4ALL library with the \texttt{mistral-7b-instruct-v} model developed by Mistral \cite{jiang2023mistral} was implemented due to its good performance in grammatical correction in Spanish and its moderate size. This model allows for grammar correction, improved cohesion, and style adaptation of the poem while maintaining the desired stylistic coherence. Additionally, the model’s \textit{prompt} can be modified to implement the desired form of address (tú, usted, vos, os, le, te, etc.). Although the language model was used as an automatic corrector, the final result was reviewed manually to ensure that it maintained the emotional and poetic charge sought in each generation.

\subsection{Potential Implementation Errors and Their Solutions}

Although the proposed algorithm is functional, it may encounter errors that cause it to enter an infinite loop (stationary states), particularly when it generates a series of "uncommon" words (i.e., words whose list of potential subsequent words contains few or no elements). This can lead to repeated cycles of the same words until the maximum requested word count is reached. 

To avoid these situations, several measures can be implemented:

1) Different versions of the same ROI can be used, creating variations in the color distribution. Additionally, color inversion can be applied, which reverses the probability direction.

2) In the case of a null options vector, force the selection of a synonym to break the cycle.

3) If no synonyms are available, use the entire corpus as the options vector, ensuring that the algorithm can continue with a valid word.

\subsection{Concrete Form}

Finally, an additional aesthetic step was implemented to give the poem the visual format of concrete poetry. This step organizes the text into various shapes that configure the screen space artistically, enhancing the visual structure of the poem. The function used includes 15 different layout schemes, and the final form selection depends on the probability value of the last word generated in the poem. Thus, the poem is expressed not only in words but also in a unique visual composition that amplifies its poetic and aesthetic impact.

\section{The Harvest of Stars: Results}

Below is a selection of poems generated by the algorithm using the gamma-ray data from the specified region.

It is important to note that the algorithm was run hundreds of times, and only the outputs that the author deemed most effective for conveying their intended emotions were selected. This selective process emphasizes the essential role of human judgment in curating and refining algorithmically generated poetry. Although the algorithm itself lacks emotional understanding, this careful curation of results enables a more authentic and resonant expression of the author’s inner world. Thus, the final selection not only showcases the algorithm’s capabilities but also reflects the author’s desire to craft a narrative that connects meaningfully with readers.

The emotional character of the poems relies entirely on the corpus used to generate them. In the author’s case, the corpus consists of a series of poems and letters written during an extended stay in a mental health facility. As a result, the poems presented here may contain themes that are sensitive for some individuals. Each poem is accompanied by its translation into English.

\subsection{First poem}

\textbf{Original version:}
\begin{verse}
Veo tus heridas me consume Cada cosa por su inocencia \\
la humanidad Mostrar mi gran y herida alguna hora jugar\\
alguien desde su inocencia está\\
\end{verse}
%\vspace{0.3cm}
\textbf{Corrected version:}
\begin{verse}
Veo tus heridas, me consumen. \\
Cada cosa por tu inocencia. \\
La humanidad muestra mi gran herida, \\
algún juego en horas. \\
Alguien desde tu inocencia está. \\
\end{verse}
%\vspace{0.3cm}
\textbf{Translated version:}
\begin{verse}
I see your wounds, they consume me.\\
Every single thing for your innocence.\\
Humanity reveals my deep wound,\\
a game played in passing hours.\\
Someone dwells within your innocence.\\
\end{verse}

\subsection{Second poem}
\textbf{Original version:}
\begin{verse}
Sombra Y además vas torcido y herida alguna nave sin\\
su casco gris ceniza te evadir nave triste Adiós Te evitaré\\
y herida dejada pasado \\
\end{verse}
%\vspace{0.3cm}
\textbf{Corrected version:}
\begin{verse}
En la sombra, yaces torcida y herida. \\
Nave sin tu casco de ceniza gris, \\
te evadiré. Triste nave, adiós; \\
te evitaré y las heridas que dejaste atrás.\\
\end{verse}
%\vspace{0.3cm}
\textbf{Translated version:}
\begin{verse}
In the shadows, you lie twisted and wounded.\\
A ship without your hull of gray ash,\\
I will evade you. Sad ship, farewell;\\
I will avoid you and the wounds you left behind.\\
\end{verse}

\subsection{Third poem}
\textbf{Original version:}
\begin{verse}
Forjado igualmente vivir todo por cumplir Y uno nuevo Y\\
volver Ser mi torre de aire Es tarde ya A\\
quién Qué rabia y amor\\
\end{verse}
%\vspace{0.3cm}
\textbf{Corrected version:}
\begin{verse}
Forjado igualmente \\
el vivir todo por cumplir,\\
y en un nuevo ser, mi torre de aire.\\
Es tarde ya; para quién rabia de amor.\\
\end{verse}
%\vspace{0.3cm}
\textbf{Translated version:}
\begin{verse}
Equally forged,\\
living fully to fulfill everything,\\
and in a new self, my tower of air.\\
It is late now; for whom rages with love.\\
\end{verse}

\subsection{Fourth poem}
\textbf{Original version:}
\begin{verse}
Cromática romperlo todo exceso contrarrestar desearía escenario sumaron producir ganaste\\
Uh oh-oh oh Te evitaré Conmigo siempre entrelazados ir con\\
\end{verse}
\textbf{Corrected version:}
\begin{verbatim}
Cromática,
         rompe
             todo
                exceso
                     que
                       deseas
                            contrarrestar;
En
 escenario
         sumarnos
                para
                   producir,
                           lo
                            ganaste.
Uh
 oh-oh,
      te
       evitaré.
Conmigo
      siempre
            entrelazada
                      irás.
\end{verbatim}
\textbf{Translated version:}
\begin{verse}
Chromatic, break every excess you wish to counteract;\\
On stage, joining together to produce,\\
you won.\\
Uh oh-oh, I will avoid you. With me always\\
intertwined you will go.\\
\end{verse}

\subsection{fifth poem}
\textbf{Original version:}
\begin{verse}
Lloro Por paisajes Donde yacen como queríamos apoyaré en agonía\\
la humanidad Mostrar subir de nivel acaparándolo despertarme y herida\\
alguna nave sin un sideral.\\
\end{verse}
\vspace{6cm}
\textbf{Corrected version:}
\begin{center}
                \textbf{Yo \\
                lloro \\
                por \\
                los \\
                paisajes \\
                donde \\
                yacen, \\
                como \\
                queríamos \\
                apoyarnos. \\
                En \\
                mi \\
                agonía, \\
                a \\
                la \\
                humanidad} \\
\textit{Yo lloro por los paisajes donde yacen, como queríamos apoyarnos. En mi agonía, a la \textit{humanidad} mostraré cómo subir al siguiente nivel, acaparando el despertar y la herida, alguna nave sin un sideral.} \\
                \textbf{cómo \\
                subir \\
                al \\
                siguiente \\
                nivel, \\
                acaparando \\
                el \\
                despertar \\
                y \\
                la \\
                herida, \\
                alguna \\
                nave \\
                sin \\
                un \\
                sideral. } \\
\end{center}
\textbf{Translated version:}
\begin{verse}
I cry for the landscapes where they lie,\\
as we wished to lean on each other.\\
In my agony, I will show humanity how to rise to the next level,\\
seizing the awakening and the wound,\\
a ship without a sidereal.\\
\end{verse}

\subsection{Sixth poem}
\textbf{Original version:}
\begin{verse}
Podrás lograr todo por su don Y volver Ser mi\\
torre darnos cuenta nos detestan organizar exterior Y uno del\\
perdón Pero yo estaré siempre\\
\end{verse}
\textbf{Corrected version:}
\begin{verse}
Podrás lograrlo todo gracias a tu don y podrás regresar. \\
Seré tu torre, haré que entendamos que nos odian, \\
organizaremos el exterior y seremos uno con el perdón.\\
Pero yo estaré siempre.\\
\end{verse}
\textbf{Translated version:}
\begin{verse}
You will be able to achieve everything thanks to your gift, and you will be able to return.\\
I will be your tower, I will make us understand that they hate us,\\
we will organize the outside and become one with forgiveness.\\
But I will always be here.\\
\end{verse}

\subsection{Seventh poem}
\textbf{Original version:}
\begin{verse}
Intentaba en cambiarme caerias Amplitud dulce anarquía común lago regresa\\
a guardarte marcharse perdurar sin su inocencia la bondad Estaba\\
en agonía conducen habilidad sentías\\
\end{verse}
\textbf{Corrected version:}
\begin{verse}
Intentaba cambiar mi situación, caíais; \\
amplitud dulce, anarquía común, el lago vuelve. \\ 
A guardaros, me marché; \\
perduraba sin tu inocencia, la bondad. \\
Estaba en agonía, conducían habilidad; sentías.\\
\end{verse}
\textbf{Translated version:}
\begin{verse}
I tried to change my situation, you fell;\\
sweet amplitude, common anarchy, the lake returns.\\
To protect you, I left;\\
enduring without your innocence, kindness.\\
I was in agony, skill was leading; you felt.\\
\end{verse}
\subsection{Eighth poem}
\textbf{Original version:}
\begin{verse}
Intentando clavar Mataría dominar \\
La distancia Haceme sería incapaz internamente
\end{verse}
\textbf{Corrected version:}
\begin{verbatim}
          Intentando
            clavarme,
              mataría
              para
              dominar
            la
          distancia
         que
       me
      hace
      ser
       incapaz
         internamente.
\end{verbatim}

\textbf{Translated version:}
\begin{verse}
Trying to pierce me, I would kill to dominate\\
the distance that makes me internally incapable.\\
\end{verse}

\subsection{Ninth poem}
\textbf{Original version:}
\begin{verse}
Aceptarlo acercar moriré próximo dada florecer de vuelo y\\
 herida alguna nave sin un efecto cómo te mueves calmar\\
  sin Sonidos extraños De neón sus\\
\end{verse}
\textbf{Corrected version:}
\begin{verse}
Acercarme a aceptar lo que me dará la muerte, \\
cerca del lugar donde florece el otro vuelo; \\
herido por algún navío en un efímero efecto. \\
¿Cómo te moverás para calmarme? \\
Sin sonidos extraños de néon.\\
\end{verse}
\textbf{Translated version:}
\begin{verse}
To draw close to accepting what will bring me death,\\
near the place where another flight blooms;\\
wounded by some vessel in a fleeting effect.\\
How will you move to calm me?\\
With no strange neon sounds.\\
\end{verse}

\subsection{Tenth poem}
\textbf{Original version:}
\begin{verse}
Hablamos tregua te cumpla Agitan proporción \\
cifra en agonía sentido estrujar alma\\
a guardarte detienes río cristalinos conectar \\
dios tocar Escondí publicación en serie en crecer más tarde
\end{verse}
\textbf{Corrected version:}
\begin{verbatim}
         Hablamos
                sin
                       tregua
cumplida,
                                 agitando
       la
         proporción
                      que
cifra
nuestra
                   agonía.
                               Sentimos
                 el
                                        estrujar
                                      del
                                    alma,
                                cuando
                   nos
                        detenemos
                                a
              contemplar
                          tu
                              río
                                        cristalino
                 y
                    conectar
                               con
                            lo
      divino.
               Tocándonos
        a
                             escondidas,
                       publicándonos
                            en
  serie,
                             y
                                creceremos
                                    más
                                    tarde.
\end{verbatim}
\textbf{Text without scattering:}
\begin{verse}
Hablamos sin tregua cumplida,\\
agitando la proporción que cifra nuestra agonía.\\
Sentimos el estrujar del alma, cuando nos detenemos\\
a contemplar tu río cristalino y conectar con lo divino. \\
Tocándonos a escondidas, publicándonos en serie,\\
y creceremos más tarde.
\end{verse}
\textbf{Translated version:}
\begin{verse}
We speak without a truce fulfilled,\\
stirring the balance that encodes our agony.\\
We feel the soul’s squeeze when we stop\\
to contemplate your crystalline river and connect with the divine.
Touching each other in secret, publishing ourselves in series,
and we will grow later.
\end{verse}
\section{The Last Light: Conclusions}
This work demonstrates the viability of using text generation algorithms for poetry by blending literary aesthetics with data processing techniques, including gamma-ray source mapping. Through the development of an algorithm that leverages a customized corpus—comprising poems, personal letters, and popular songs—poems were created that both echo the author’s voice and embody unique artistic expression. However, the algorithm lacks emotional understanding, depending solely on the quality and diversity of the input. Some limitations, such as looping on uncommon words, can be mitigated by strategies like expanding the gamma-ray source map or substituting synonyms to maintain coherence. Finally, aesthetic choices, such as varied formatting and color distributions, deepen the reader’s experience and underscore the interplay between poetry and visual art. Consequently, this work not only opens the door to new forms of literary creation but also raises questions about the future of algorithmically generated poetry, its reception by the public, and its potential as a tool for artistic expression.

\textit{Thank you for reading my poems, a part of my soul.}

\section{Where the stars speak}
In this repository, you will find an implementation of the algorithm in Python using Google Colab. The corpus includes poems by the author and a selection of texts and songs that have accompanied them throughout their life. 

\hyperlink{https://github.com/physicsIS/Physics-in-Arts/tree/main/Poetry}{Link to Github}

\addcontentsline{none}{section}{References}
\printbibliography

\end{document}